\documentclass[times,review,10pt]{elsarticle}

\usepackage{amssymb}
\usepackage{algorithm}
\usepackage{algorithmic}
\usepackage{graphicx}

\usepackage{lineno} 
\usepackage{amsmath}

\usepackage{float}
\usepackage{tabularx,array,booktabs}

\usepackage{booktabs}
\usepackage{multirow}
\usepackage{longtable} 
\usepackage{makecell}

\usepackage{xcolor}

\journal{Computerized Medical Imaging and Graphics}

\begin{document}
\begin{frontmatter}



\title{Fully Guided Neural Schr\"odinger Bridge for Brain MR Image Synthesis}


\cortext[cor1]{Corresponding author}

\author[1]{Hanyeol Yang}
\ead{dfgtyrui@hanyang.ac.kr}
\author[3]{Sunggyu Kim}
\author[2]{Mi Kyung Kim}
\author[1]{Yongseon Yoo}
\author[2]{Yu-Mi Kim}
\author[4]{Min-Ho Shin}
\author[5]{Insung Chung}
\author[6]{Sang Baek Koh}
\author[7]{Hyeon Chang Kim}
\author[1,3]{Jong-Min Lee\corref{cor1}}
\ead{ljm@hanyang.ac.kr}

\affiliation[1]{organization={Hanyang University, Department of Artificial Intelligence},
            addressline={222 Wangsimni-ro, Seongdong-gu}, 
            city={Seoul},
            postcode={04763}, 
            state={Seoul},
            country={Republic of Korea}}
            
\affiliation[2]{organization={Hanyang University College of Medicine, Department of Preventive Medicine},
            addressline={222 Wangsimni-ro, Seongdong-gu}, 
            city={Seoul},
            postcode={04763}, 
            state={Seoul},
            country={Republic of Korea}}
            
\affiliation[3]{organization={Hanyang University, Department of Biomedical Engineering},
            addressline={222 Wangsimni-ro, Seongdong-gu}, 
            city={Seoul},
            postcode={04763}, 
            state={Seoul},
            country={Republic of Korea}}

\affiliation[4]{organization={Chonnam National University Medical School, Department of Preventive Medicine},
            addressline={160, Baekseo-ro, Dong-gu}, 
            city={Gwangju},
            postcode={61469}, 
            state={Gwangju},
            country={Republic of Korea}}
            
\affiliation[5]{organization={Keimyung University School of Medicine, Department of Occupational and Environmental Medicine},
            addressline={1095 Dalgubeoldaero, Dalseo-gu}, 
            city={Daegu},
            postcode={42601}, 
            state={Daegu},
            country={Republic of Korea}}
            
\affiliation[6]{organization={Yonsei Wonju College of Medicine, Department of Preventive Medicine and Institute of Occupational Medicine},
            addressline={162 Ilsan-Dong}, 
            city={Wonju},
            postcode={220-701}, 
            state={Wonju},
            country={Republic of Korea}}
            
\affiliation[7]{organization={Yonsei University College of Medicine, Department of Preventive Medicine},
            addressline={50-1, Yonsei-Ro, Seodaemun-gu}, 
            city={Seoul},
            postcode={03722}, 
            state={Seoul},
            country={Republic of Korea}}

\begin{abstract}
Multi-modal brain MRI provides essential complementary information for clinical diagnosis. However, acquiring all modalities in practice is often constrained by time and cost. To address this, various methods have been proposed to generate missing modalities from available ones. Existing approaches can be broadly categorized into two types: paired and unpaired methods. While paired methods achieve high synthesis accuracy, obtaining large-scale paired datasets is typically impractical. In contrast, unpaired methods, though more scalable, often fail to preserve critical anatomical features, such as lesions. In this paper, we propose Fully Guided Schr\"odinger Bridge (FGSB), a novel framework designed to overcome these limitations by enabling high-fidelity generation with extremely limited paired data. When lesion-specific information, such as expert annotations or segmentation masks, is available, FGSB preserves clinically relevant lesions during missing modality synthesis. Our model comprises two stages: (1) a generation stage that iteratively refines synthetic images using paired source images and Gaussian noise, and (2) a training stage that learns optimal transformation pathways by modeling intermediate states to ensure consistent, high-fidelity synthesis. Experimental results across multiple datasets demonstrate that FGSB achieves reliable synthesis performance across diverse imaging resolutions and data acquisition environments. In addition, incorporating lesion-specific priors further enhances the preservation of clinically relevant features.
\end{abstract}



\end{frontmatter}
\setcounter{page}{2}



\section{Introduction}
Multi-modal magnetic resonance imaging (MRI) of the brain provides complementary information about anatomy and pathology across different sequences, which is crucial for accurate diagnosis and robust segmentation of regions of interest~\cite{ZHANG2026102703,KOC2026102727}. However, acquiring all modalities for every patient is often infeasible due to constraints in time, cost, and patient compliance. To address this challenge, various methods have been proposed to synthesize missing modalities from acquired ones. Recent advances in medical image synthesis have been driven by deep learning-based frameworks such as Generative Adversarial Networks (GANs)~\cite{LUO202163,10124026}, diffusion models~\cite{10.1007/978-3-031-43999-5_38}, and Vision Transformers~\cite{9758823,9774943}. Medical image synthesis, which can be viewed as a specific application of image-to-image translation, is broadly categorized into two approaches~\cite{10.5555/3540261.3540412}: paired learning and unpaired learning. Paired learning requires aligned source-target image pairs and achieves high fidelity, but depends on scarce paired datasets. In contrast, unpaired learning works with unaligned images and enables larger-scale training, but often fails to preserve fine-grained anatomical details such as lesions. In addition, unpaired learning methods typically rely on additional networks~\cite{8653423}, such as those enforcing cycle-consistency~\cite{8237506}.

Among paired methods, MT-Net~\cite{10158035} incorporates MAE~\cite{9879206}-based pre-training using a Vision Transformer (ViT)~\cite{dosovitskiy2021an} encoder to overcome the scarcity of paired data. However, effective MAE-based pre-training requires a large amount of data~\cite{10.1007/978-3-031-72390-2_32}, and the available number of subjects for both pre-training and fine-tuning may be substantially limited in practice.

Acquiring sufficient paired data remains challenging in practice. Consequently, unpaired learning approaches have been widely adopted in medical image synthesis. However, many unpaired learning methods rely on cycle-consistency to preserve anatomical structures or lesions~\cite{choi2024ct,YURT2021101944}. Cycle-consistency-based approaches introduce additional model complexity and may not reliably preserve fine-grained anatomical structures~\cite{10.1007/978-3-031-43999-5_6,Gong2024}.

Furthermore, GAN-based models characterize the target modality distribution using implicit adversarial learning rather than explicit likelihood estimation. This indirect modeling approach can introduce training instabilities that manifest as premature convergence and mode collapse~\cite{10.5555/3540261.3540933}.

SynDiff~\cite{10167641} demonstrates a promising application of diffusion models~\cite{NEURIPS2020_4c5bcfec} in medical image synthesis. However, the absence of an explicit mechanism to ensure consistency across intermediate states may compromise the stability of its generation process. Additionally, the dependency on cycle-consistency in unpaired learning approaches remains a fundamental limitation.

To address these limitations, we propose FGSB, a neural Schr\"odinger bridge-based architecture for medical image synthesis. FGSB generates a corresponding target image given a source image and gradually refines the target image using Gaussian noise. Unlike standard diffusion models, FGSB uses a small number of time steps and employs mutual information loss to maintain consistency across the intermediate samples generated during the process. Consequently, without any pre-training, FGSB achieves performance comparable to competing methods trained on larger datasets across diverse experimental scenarios, including datasets with a limited number of subjects and those with a restricted number of axial slices per subject.

\section{Related Works}
\subsection{Neural Schr\"odinger Bridges}
The Schr\"odinger bridge (SB) finds optimal transport trajectories between arbitrary source and target distributions by progressively sampling intermediate states, and has recently been applied to diffusion-based image translation~\cite{10203692,su2022dual,pmlr-v202-liu23ai}. Unlike traditional approaches limited by Gaussian assumptions, UNSB~\cite{kima2024unpaired} formulates the SB as an adversarial learning problem, enabling efficient learning in high-dimensional spaces.

The process decomposes as a Markov chain where each intermediate sample ${x_t}$ is characterized by~\cite{tong2024improving}:

\begin{equation}
p(x_t|x_A, x_B) = \mathcal{N}(x_t|tx_B + (1-t)x_A, t(1-t)\tau\mathbf{I})
\end{equation}

UNSB demonstrates that the Schr\"odinger Bridge can be effectively represented as a composition of adversarial learning and a Markov chain:

\begin{equation}
\begin{aligned}
p(\{x_{t_n}\}) = p(x_{t_N}\vert x_{t_{N-1}})p(x_{t_{N-1}}\vert x_{t_{N-2}})
\\ \cdots p(x_{t_1}\vert x_{A})p(x_{A})
\end{aligned}
\end{equation}

Eq. (2) demonstrates that the generation can be decomposed into a sequence of conditional distributions. Given the source image $x_A$, we can sequentially sample intermediate states ${x_{t_i}}$ at each time step. This iterative sampling process culminates in the final predicted target image $\hat{x}_{B_T}$, which approximates the  target distribution.

\begin{equation}
q_{\phi_i}(x_{t_i}, x_B) := q_{\phi_i}(x_B\vert x_{t_i})p(x_{t_i})
\end{equation}
\begin{equation}
q_{\phi_i}(x_B) := \mathbb{E}_{p(x_{t_i})}[q_{\phi_i}(x_B\vert x_{t_i})]
\end{equation}

Eq. (3) defines the joint distribution between the intermediate state $x_{t_i}$ and the target image $x_B$, where $q_{\phi_i}$ predicts the target image given an intermediate state. Eq. (4) expresses the marginal distribution of generated target images by integrating over all possible intermediate states. This formulation provides a probabilistic framework for modeling the relationship between intermediate states and the target image. 

\begin{equation}
\begin{aligned}
\min_{\phi_i} \mathcal{L}_{SB}(\phi_i, t_i) &:= \mathbb{E}_{q_{\phi_i}(x_{t_i}, x_B)}[\|x_{t_i} - x_B\|^2] - 2\tau(1-t_i)H(q_{\phi_i}(x_{t_i}, x_B)) \\
&\text{s.t.} \, \mathcal{L}_{adv}(\phi_i, t_i) := D_{KL}(q_{\phi_i}(x_B)\|p(x_B)) = 0
\end{aligned}
\end{equation}

$q_{\phi_i}$ is a generator that produces intermediate samples along the trajectory to the target image $x_B$. $q_{\phi_i}$ is parameterized by a neural network. Eq. (5) demonstrates that $\mathcal{L}_{adv}$ serves as a crucial learning condition for SB. UNSB utilized an enhanced discriminator architecture, which is particularly justified given the constraints of finite sampling and the curse of dimensionality encountered in mid-time sampling processes.

\begin{table}[!h]
\centering
\scalebox{0.85}{%
\resizebox{0.8\textwidth}{!}{%
\label{tab:notation}
\begin{tabular}{ll}
\hline
\renewcommand{\arraystretch}{0.6}
Symbol & Description \\
\hline
$x_A$ & The source domain image (T1w image) \\
$x_B$ & The target domain image (e.g., T2w, FLAIR, etc.) \\
$q_\phi$ & The generator network that maps source to target domain \\
$D$ & The discriminator network \\
$E$ & The mutual information estimator \\
$NFE$ & Number of Function Evaluations (maximum of time steps) \\
$T$ & Randomly sampled time step from $\{0, \ldots, NFE\}$ \\
$i$ & Intermediate time step index ($i \in \{0, \ldots, T\}$) \\
$x_{t_i}$ & Input intermediate sample at timestep $i$ \\
$\hat{x}_{B_i}$ & Generated intermediate sample at time step $i$ \\
$\hat{x}_{B_T}$ & Generated target image at sampled time step $T$ \\
$\tau_{t_i}$ & Gaussian noise variance (stochasticity parameter) \\
$\epsilon_{t_{i}}$ & ${\sim}\, \tau_{t_i} \cdot \mathcal{N}(0, I)$ \\
$s_{t_i}$ & Predefined interpolation value at time step $i$ \\
$x_{prior}$ & The binary map for lesion-specific guidance\\
\hline
\end{tabular}
}
}
\caption{Key symbols and their descriptions.}
\label{notations}
\end{table}

UNSB consists of two stages: Generation and Training. The generation stage produces intermediate samples by combining three components: the input image, the previous generator sample, and Gaussian noise with predefined variance. The training stage uses adversarial loss and patchNCE~\cite{10.1007/978-3-030-58545-7_19} loss to translate the output to the target modality while preserving anatomical details.

Despite its innovative approach, direct application of the UNSB framework presents several limitations. While UNSB employs unpaired learning to generate subsequent time step samples solely through network input and output without any target modality information, the unpaired learning strategy in UNSB potentially compromises critical lesion information present in the source image. To address this limitation, we incorporate the source modality into both the generation and training stages, and the target modality information exclusively into the training stage. This paired learning paradigm enables the use of reconstruction loss~\cite{8100115} and facilitates the integration of additional prior information, such as expert annotations or segmentation labels.

\begin{algorithm}[H]
\caption{Our framework workflow}
\small
\begin{algorithmic}[1]
\setlength{\itemsep}{-5pt} 
\STATE {\bf Input:} Source domain image $x_A$, target domain image $x_B$, $NFE$, noise variance $\tau$, interpolation variable $s_{t_i}$
\STATE {\bf Output:} Translated image $\hat{x}_{B_T}$ 
\STATE {\bf Parameters:} Generator $q_\phi$, discriminator $D$, MI estimator $E$
\STATE
\STATE {\bf Training Stage:}
\STATE Sample random time step $T \in \{0, \ldots, NFE\}$
\STATE Calculate $\hat{x}_{B_T}$ (see Generation Stage) \COMMENT{Generate target image}
\STATE Update $D \leftarrow L_{adv}^D$, $E \leftarrow L_{SB}^E$, $q_{\phi} \leftarrow L_{FGSB}$
\STATE
\STATE {\bf Generation Stage:}
\STATE \quad $x_{t_0} \leftarrow x_A$, $i = 0$ {\bf to} $T+1$
\STATE \quad \quad $\hat{x}_{B_i} \leftarrow q_\phi(x_{t_i}, x_A, i, z)$ \COMMENT{Generate intermediate samples}
\STATE \quad \quad {\bf Training:} $x_{t_{i+1}} \leftarrow (1-s_{t_{i}})x_A + s_{t_{i}}\hat{x}_{B_i} + \tau_{t_i} \cdot \mathcal{N}(0, I)$
\STATE \quad \quad {\bf Inference:} $x_{t_{i+1}} \leftarrow (1-s_{t_{i}})x_A + s_{t_{i}}\hat{x}_{B_i} + \tau_{t_i} \cdot \mathcal{N}(0, I)$
\STATE \quad {\bf Return:} $\hat{x}_{B_T}$
\end{algorithmic}
\label{learning_process}
\end{algorithm}

\begin{figure*}[!h]
\centerline{\includegraphics[width=\linewidth]{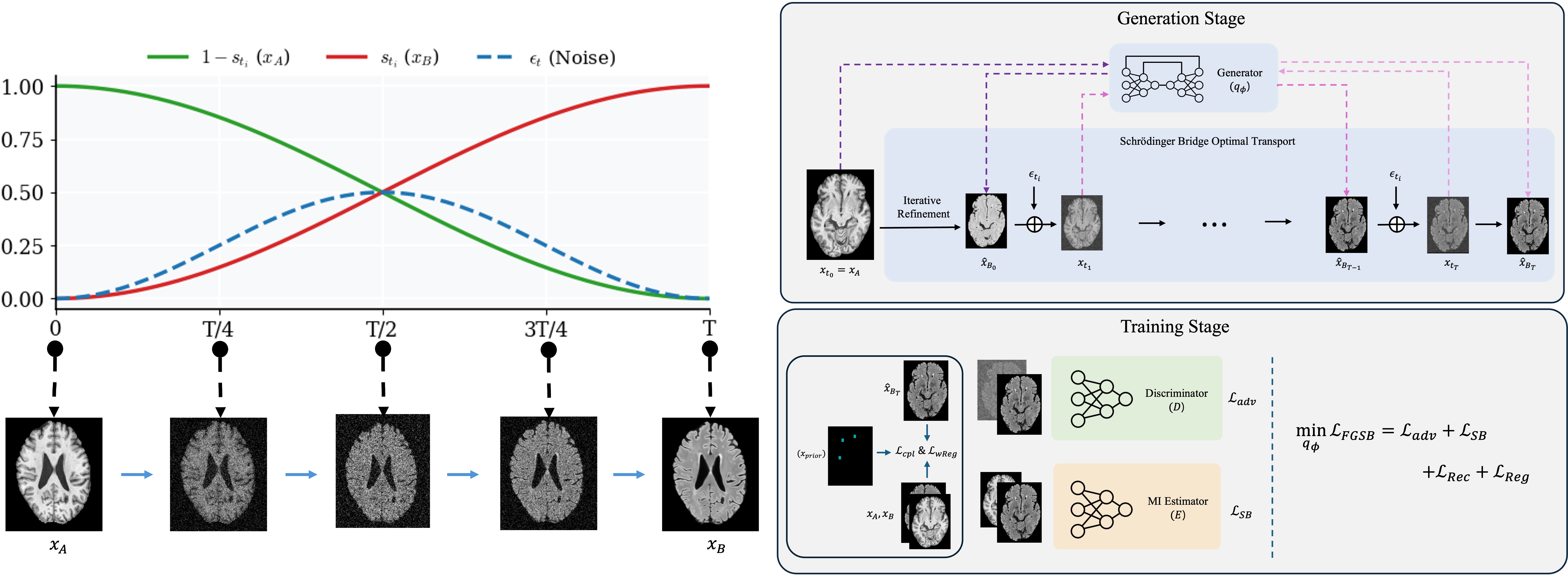}}
\caption{\textbf{Overview of the proposed Fully Guided Schr\"odinger Bridge (FGSB) framework.} Left: Conceptual illustration of Schr\"odinger Bridge optimal transport between source $\{x_A\}$ and target $\{x_B\}$ distributions through generated intermediate samples $\{\hat{x}_B\}$. Right: Generation Stage (top): The generator $q_\phi$ iteratively refines the intermediate samples by combining the generated intermediate sample $\hat{x}_{B_i}$, Gaussian noise $\epsilon_{t_{i}}$, and paired images $x_A$. Training Stage (bottom): At each training step, a time step $T$ is uniformly sampled from $\{0, \ldots, NFE\}$ to generate the target image $\hat{x}_{B_T}$ for optimization, enabling the model to learn the entire trajectory across training iterations. The framework optimizes generator $q_\phi$ (yellow) with self-supervised discriminator $D$ (green) for $\mathcal{L}_{adv}$, MI estimator $E$ (orange) for $\mathcal{L}_{SB}$, and identity loss $\mathcal{L}_{idt}$ for inference stability. Additional losses ($\mathcal{L}_{Rec}$, $\mathcal{L}_{Reg}$) incorporate optional prior information ($x_{prior}$). The final objective $\mathcal{L}_{FGSB}$ combines all terms for high-fidelity synthesis with limited data.}
\label{Figure_1}
\end{figure*}

\section{Method} 
\subsection{Overview}
\begin{figure*}[!h]
\centerline{\includegraphics[width=\textwidth]{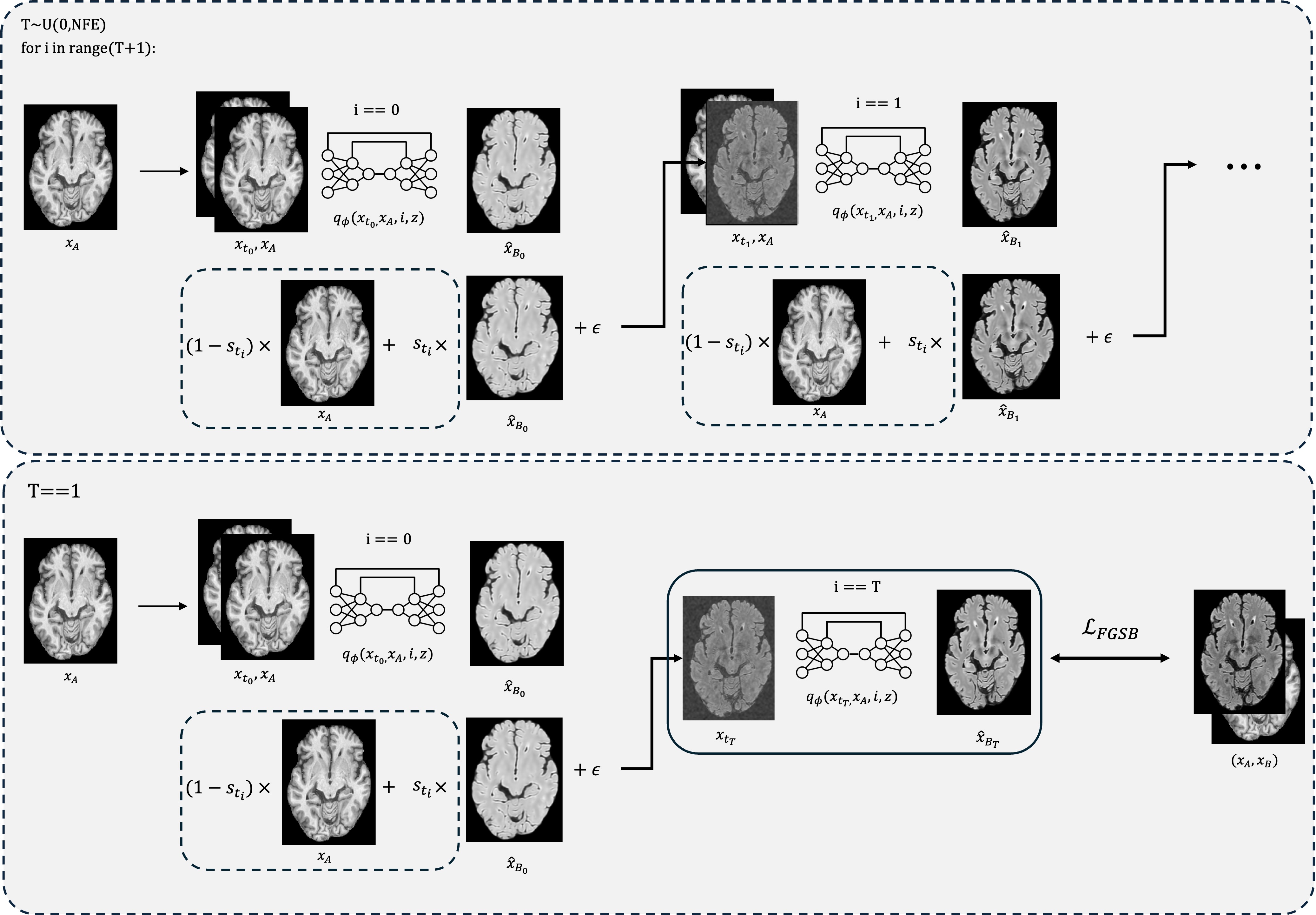}}
\caption{\textbf{Detailed illustration of generation and training stages with concrete examples.} Top: General iterative generation stage. Starting from source image $x_A$, the generator $q_\phi$ produces intermediate sample $\hat{x}_{B_0}$ at $i=0$. The next input intermediate sample $x_{t_1}$ is computed by weighted combination: $(1-s_{t_i}) \times x_A + s_{t_i} \times \hat{x}_{B_0} + \epsilon_{t_i}$, where $s_{t_i}$ controls interpolation between source image $x_A$ and generated intermediate sample $\hat{x}_{B_0}$, and $\epsilon_{t_{i}}$ adds Gaussian noise. This iterative refinement continues for subsequent steps ($i=1, 2, \ldots$). Bottom: Training stage example when time step $T=1$ is randomly sampled. The generation stage produces generated target image $\hat{x}_{B_T}$ at the sampled time step, which is then used along with the paired images $(x_A, x_B)$ to compute the composite loss $\mathcal{L}_{FGSB}$ for optimization. By randomly sampling different time steps across all dataset iteration, the model learns the entire trajectory of intermediate distributions.}
\label{Figure_2}
\end{figure*}

Our proposed Fully Guided Schr\"odinger Bridge (FGSB) framework (Fig. \ref{Figure_1}, \ref{Figure_2}, and Algorithm \ref{learning_process}) comprises two main stages: a generation stage and a training stage. The generation and training stages are executed sequentially. When a specific time step $T$ is sampled, only the intermediate sample ($\hat{x}_{B_T}$) is generated and subsequently used in the training stage. Here, $i$ denotes the intermediate time step index along the generation trajectory leading up to the sampled $T$, where $i = 0,\ldots,T$. Starting from $x_{t_0}=x_A$ at $i=0$, the generator produces intermediate sample $\hat{x}_{B_i}$ and updates the input intermediate sample $x_{t_{i+1}}$ iteratively until reaching the final output $\hat{x}_{B_T}$.

\subsection{Generation stage}
The generation stage is responsible for producing intermediate samples and ultimately generating the synthetic target image. This process is designed to be both temporally guided and stochastic, enabling iterative refinement across multiple steps, as detailed below.

\subsubsection{Initialization ($x_{t_0}$)}
When a specific time step $T$ is sampled, the intermediate steps leading up to $T$ are indexed as $i$. At the initial time step $i=0$, the source image $x_A$ is directly passed into the generator $q_{\phi_i}$. This serves as the starting point for the synthesis toward the target image $x_B$. Therefore, the input to $q_{\phi_i}$ is denoted as: 

\begin{equation}
    x_{t_0} = x_A 
\end{equation}

where $x_A$ is the input image from the source modality.

\subsubsection{Iterative generation process}
For each intermediate time step $i \in [0, T]$, the generator $q_{\phi}$ takes the intermediate sample ${x}_{t_{i}}$ as input and produces $\hat{x}_{B_{i}}$. The next intermediate sample ${x}_{t_{i+1}}$ is then constructed by combining the source image $x_A$, the generator output $\hat{x}_{B_{i}}$, and Gaussian noise $\epsilon_{t_{i}} \sim \tau_{t_i}\mathcal{N}(0,I)$, following the Schr\"{o}dinger bridge formulation~\cite{tong2024improving}:

\begin{equation}
x_{t_{i+1}} = ((1 - s_{t_i}) \cdot x_A) + (s_{t_i} \cdot \hat{x}_{B_{i}}) + \epsilon_{t_{i}}
\end{equation}

$\tau_{t_i}$ is a scalar hyperparameter controlling the stochasticity of the entire SB trajectory, and $\hat{x}_{B_{i}}$ is the prediction of the generator $q_{\phi_i}$ from the current time step $i$. The iterative generation process enforces smooth progression toward target image $x_B$.

$s_{t_i}$ is a hyperparameter predefined for each time step to determine~\cite{pmlr-v202-liu23ai} the relative importance between the source image $x_A$ and the model output $\hat{x}_{B_{i}}$ (Fig 1, left).

\subsection{Training stage}
Given a dataset of paired images $\{(x_A, x_B)\}$, where $x_A$ denotes the source image (T1w image) and $x_B$ is the corresponding target image (e.g., T2w, FLAIR, etc.), we randomly sample a time step $T \in [0, NFE]$ during training. The generator then produces an intermediate sample $\hat{x}_{B_T}$.

The training stage aims to optimize the parameters of the generator to achieve high-fidelity synthesis aligned with the target image. This is accomplished by minimizing a composite loss function that includes adversarial $\mathcal{L}_{adv}$, reconstruction $\mathcal{L}_{Rec}$, patchNCE $\mathcal{L}_{Reg}$, and mutual information loss $\mathcal{L}_{SB}$. We detail each element of the training stage below.

We adopt a non-saturating adversarial loss to train the generator $q_\phi$ and discriminator $D$. Unlike standard GAN-based approaches that discriminate only between the final generated image and the real target, our discriminator operates on trajectory-level samples constructed according to the Schr\"{o}dinger bridge interpolation schedule. Specifically, the discriminator receives the intermediate sample $x_{t_T}$ computed from the ground-truth pair $(x_A, x_B)$ via the predefined SB schedule (denoted $x_{t_{T|AB}}$) and the paired target image $x_B$ as real samples. As fake samples, it receives the corresponding intermediate sample $x_{t_T}$ constructed from the pair $(x_A, \hat{x}_{B_T})$ and the generator prediction $\hat{x}_{B_T}$. This trajectory-aware discrimination provides a stronger training signal than endpoint-only adversarial supervision, encouraging the generator to produce intermediate samples consistent with the true modality transition throughout the entire generation stage. To stabilize discriminator training under limited data conditions, we apply a lazy R1 gradient penalty at regular intervals, computed with respect to the real target image $x_B$:

\begin{equation}
    \mathcal{L}_{R1} = \frac{\gamma}{2} \mathbb{E}_{x_B \sim p_{\text{data}}} \left[ \left\| \nabla_{x_B} D(x_B) \right\|^2 \right]
\end{equation}

\begin{equation}
    \min_{q_\phi} \mathcal{L}_{adv} =\mathbb{E}_{\hat{x}_{B_T}}[-log(D(\hat{x}_{B_T},x_{t_T}))]
\end{equation}

\begin{equation}
    \min_{D} \mathcal{L}_{adv} = \mathbb{E}_{\hat{x}_{B_T}}[-log(1-(D(\hat{x}_{B_T},x_{t_T})))] + \mathbb{E}_{x_{B}}[-log(D(x_{B},x_{t_{T|AB}}))] + \mathcal{L}_{R1}
\end{equation}

where $\gamma$ is a weighting coefficient. By computing the penalty~\cite{8653423} only periodically rather than at every training step, this lazy regularization reduces computational overhead while effectively constraining the discriminator gradient norm at real data points.

We apply reconstruction and patchNCE loss to ensure anatomical preservation. We define $F$ as a two-layer MLP network used to compute the patchNCE loss, $\mathcal{L}_{Reg}$.

\begin{equation}
\min_{q_\phi} \mathcal{L}_{Rec} = \mathbb{E}_{\hat{x}_{B_T}}[|\hat{x}_{B_T} - x_B|]
\end{equation}

\begin{equation}
\min_{q_\phi} \mathcal{L}_{Reg} = \mathbb{E}_{\hat{x}_{B_T}}[F(\hat{x}_{B_T}, x_A)] + \mathbb{E}_{\hat{x}_{B_T}}[F(\hat{x}_{B_T}, x_B)]
\end{equation}

To enforce semantic consistency across intermediate steps (Fig. 1), we incorporate a patch-wise mutual information loss, $\mathcal{L}_{SB}$, using a neural mutual information estimator~\cite{pmlr-v80-belghazi18a}. In the Schr\"{o}dinger bridge formulation, intermediate samples $x_{t_i}$ are subject to increasing uncertainty as the generation trajectory progresses, particularly when operating with a small $NFE$. Without explicit regularization, this stochasticity can cause intermediate samples to drift away from the target modality, destabilizing the generation. $\mathcal{L}_{SB}$ addresses this by acting as a trajectory-level regularizer that maintains semantic alignment between intermediate samples and the target distribution at each timestep, thereby stabilizing the bridge under low-$NFE$ conditions. The mutual information estimator $E$ and the generator are jointly optimized through an adversarial training scheme, following a min-max optimization paradigm. The mutual information estimator loss is defined as follows:

\begin{equation}
\min_{E} \mathcal{L}_{SB} = \mathbb{E}_{x_B}[-E(x_A, x_B)]
\end{equation}

The mutual information loss is defined as:

\begin{equation}
\min_{q_\phi} \mathcal{L}_{SB} = \mathbb{E}_{\hat{x}_{B_T}}[-E(x_A, \hat{x}_{B_T})]
\end{equation}

Then, the final loss $\mathcal{L}_{FGSB}$ consists of $\mathcal{L}_{adv}$, $\mathcal{L}_{SB}$, $\mathcal{L}_{Rec}$, and $\mathcal{L}_{Reg}$, where $F$ is the learnable parameter.

\begin{equation}
\begin{aligned}
\min_{q_\phi} \mathcal{L}_{FGSB} := & \mathcal{L}_{adv} + \lambda_{SB}\mathcal{L}_{SB} + \lambda_{Rec}\mathcal{L}_{Rec}  \\
& + \lambda_{Reg}\mathcal{L}_{Reg}
\end{aligned}
\end{equation}

To incorporate lesion-specific information into the synthesis, we apply a combined loss consisting of context-preserving loss~\cite{mo2019instagan} and weighted patchNCE loss (excluding the IXI experiment). The lesion prior map $x_{prior}$ is constructed based on the availability of segmentation labels for each dataset. For the MICCAI2017 WMH dataset and VALDO, expert-annotated segmentation masks are used directly as $x_{prior}$, corresponding to WMH regions and cerebral microbleeds (CMBs), respectively. For the CAVAS dataset, pseudo-annotations generated by a pre-trained WMH segmentation network are employed in place of manual labels.

\begin{equation}
\min_{q_\phi} \mathcal{L}_{cpl} = \mathbb{E}_{\hat{x}_{B_T} \odot x_{prior}}[x_{prior} \odot (\hat{x}_{B_T} - x_B)^2]
\end{equation}

\begin{figure*}[!t]
\centerline{\includegraphics[width=\linewidth]{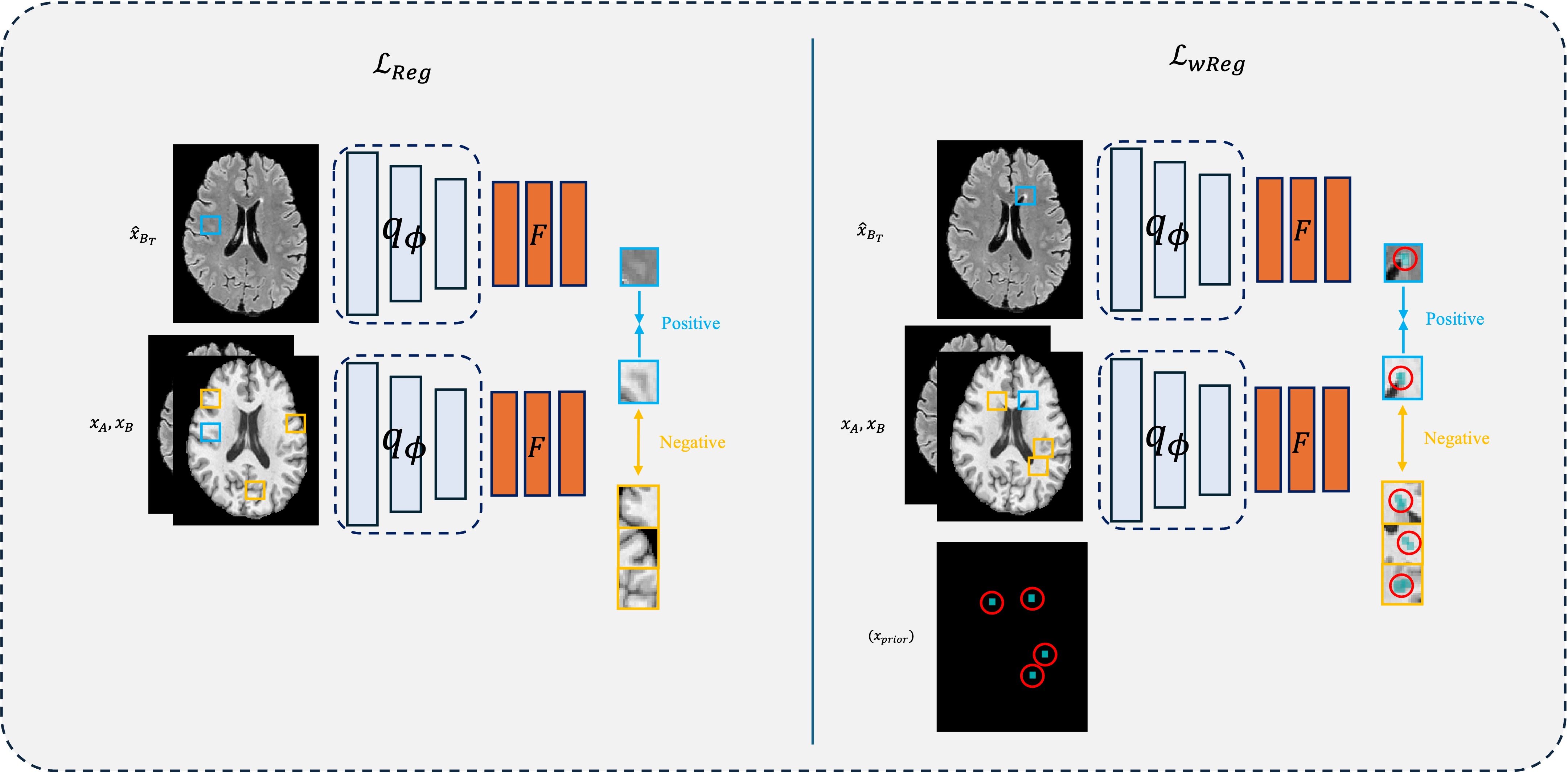}}
\caption{Comparison between standard patchNCE ($\mathcal{L}_{Reg}$) and weighted patchNCE ($\mathcal{L}_{wReg}$). Both methods extract features from the generated target image ($\hat{x}_{B_T}$) and source ($x_A$) images through multiple early layers of the generator ${q_\phi}$ (indicated by dashed boxes), which are then processed by the MLP network F for contrastive learning. Left: Standard patchNCE loss ($\mathcal{L}_{Reg}$) uniformly samples patches across all image regions (except for background) without prioritization. Right: Weighted patchNCE loss ($\mathcal{L}_{wReg}$) incorporates lesion-specific prior information ($x_{prior}$, shown in the bottom) to guide the patch sampling, prioritizing patches from clinically relevant regions such as white matter hyperintensities (indicated by red circles). This prior-guided sampling strategy enables better preservation of critical anatomical features during synthesis.}
\label{Figure_3}
\end{figure*}

The patchNCE loss computation requires a sampling of both positive and negative patches for contrastive learning. Therefore, we first utilize $x_{prior}$ to identify and select critical patches for the sampling procedure (Fig. \ref{Figure_3}).

\begin{equation}
\min_{q_\phi} \mathcal{L}_{wReg} = \mathbb{E}_{\hat{x}_{B_T}}[F(\hat{x}_{B_T}, x_A, x_{prior})] + \mathbb{E}_{\hat{x}_{B_T}}[F(\hat{x}_{B_T}, x_B, x_{prior})]
\end{equation}

The complete loss includes:

\begin{equation}
\begin{aligned}
\min_{q_\phi} \mathcal{L}_{FGSB} := & \lambda_{adv}\mathcal{L}_{adv} + \lambda_{SB}\mathcal{L}_{SB}
+ \lambda_{Rec}\mathcal{L}_{Rec} \\
& + \lambda_{cpl}\mathcal{L}_{cpl} + \lambda_{Reg}\mathcal{L}_{Reg} + \lambda_{wReg}\mathcal{L}_{wReg}.
\end{aligned}
\end{equation}

\section{Experiments}
We evaluated our framework using several datasets (IXI~\cite{braindevelopmentDatasetx2013}, BraTS2020~\cite{bakas_etal_2018}, MICCAI2017 WMH challenge dataset~\cite{8669968} and VALDO~\cite{SUDRE2024103029}). Furthermore, to assess clinical applicability, we conducted additional evaluation on an internal CAVAS dataset. All datasets were split in a subject-wise manner into non-overlapping training and test sets. All images underwent brain extraction~\cite{HDBET} to remove non-brain tissues, followed by intensity normalization to the range [-1, 1] and padding to uniform dimensions (224×224 or 256×256).

\subsection{Dataset}
\subsubsection{IXI dataset}
We used 1.5T T1-weighted (T1w) and T2-weighted (T2w) MRI data from 25 training and 10 test subjects. For each subject, 100 axial cross-sectional slices were extracted. Background-dominated slices were excluded, and all images were spatially co-registered.

\subsubsection{BraTS 2020}
We used T1w, T2w, and FLAIR from 25 low-grade glioma subjects (15 training, 10 test) across multiple cohorts. Subjects were visually inspected for consistent image contrast and tissue characteristics, with background slices excluded from 70 extracted axial slices per subject.

\subsubsection{MICCAI2017 WMH challenge dataset}
We used 3T T1w and FLAIR from 20 participants (10 training, 10 test). FLAIR was co-registered to T1w, background-dominated slices were excluded, yielding approximately 300 axial slices in both the training and test sets.

\subsubsection{VALDO}
T2* image synthesis from T1w was evaluated using approximately 400 and 180 axial slices from 19 and 8 subjects for training and test, respectively. Notably, this dataset contains a very limited number of axial slices per subject, and T2* map synthesis from T1w represents a considerably more challenging task than FLAIR synthesis.

\begin{figure*}[h]
\centering
\includegraphics[width=0.99\textwidth]{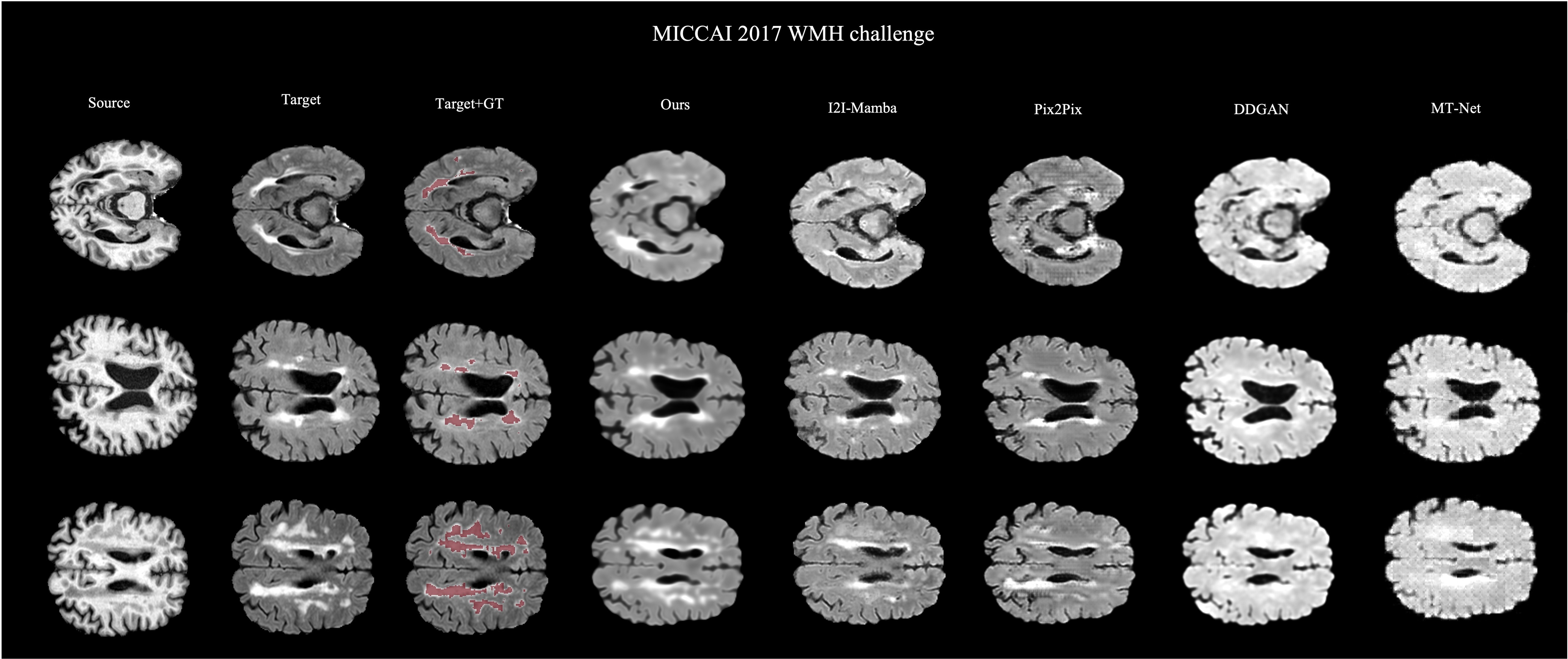}
\caption{Qualitative comparison for FLAIR synthesis on MICCAI2017 WMH challenge between our method and other methods.}
\label{IXI_FLIAR_visualization1}
\end{figure*}

\begin{figure*}[h]
\centering
\includegraphics[width=0.99\textwidth]{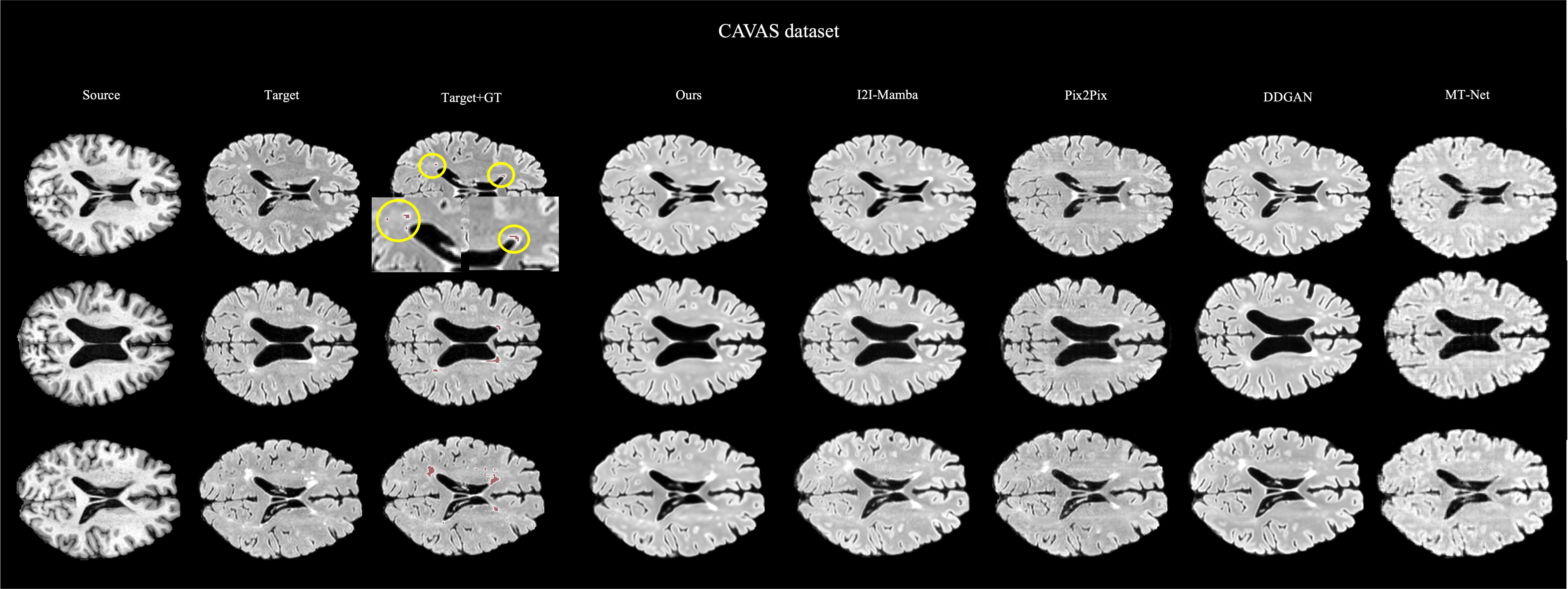}
\caption{Qualitative comparison for FLAIR synthesis on CAVAS dataset between our method and other methods.}
\label{IXI_FLIAR_visualization2}
\end{figure*}

\subsubsection{CAVAS dataset}
The CAVAS dataset is a non-public, internal dataset collected from South Korea under the approval of the Institutional Review Board (IRB). It contains 3T T1w and FLAIR images. After co-registration and background removal of 50 axial slices, we used 10 subjects for training and 24 for evaluation.

\subsection{Evaluation Metrics}
We employed multiple evaluation metrics to assess the quality of generated images. Image reconstruction fidelity was measured using Peak Signal-to-Noise Ratio (PSNR), Structural Similarity Index (SSIM), and Normalized Root Mean Square Error (NRMSE).

For FLAIR synthesis evaluation, we applied a pre-trained segmentation network~\cite{PARK2021118140} on both real and synthetic FLAIR images, comparing results with ground truth (MICCAI2017) or pseudo annotations (CAVAS dataset).

\subsection{Comparison Methods}
We compared our method against cGAN (Pix2Pix)~\cite{8100115}, DDGAN~\cite{xiao2022tackling}, I2I-Mamba~\cite{atli2025i2imambamultimodalmedicalimage} and MT-Net~\cite{10158035}. For reproducibility and fair comparison, all competing methods were reimplemented using the officially released code provided by the original authors. Training configurations, preprocessing pipelines, and hyperparameter settings were strictly followed as described in the respective publications, including learning rate, batch size, optimizer, and number of training epochs. All models were evaluated under identical experimental conditions, including the same data splits, input resolutions, and evaluation metrics.

DDGAN: a denoising diffusion GAN that combines the iterative refinement of diffusion models with adversarial training to accelerate sampling. In our experiments, DDGAN was employed under a paired training paradigm, equivalent to the SynDiff implementation with the cycle-consistency components removed, serving as a diffusion-based paired learning baseline.

MT-Net: a paired learning method that incorporates MAE-based pre-training using a Vision Transformer (ViT) encoder to address the scarcity of paired training data. The encoder is pre-trained on large-scale unlabeled data to learn generalizable image representations, which are subsequently fine-tuned for cross-modality synthesis. For pre-training, the following configurations were adopted: IXI: approximately 311 subjects from the IXI dataset; BraTS2020: officially released pre-trained weights were used; VALDO: pre-training was performed using the same data as fine-tuning; MICCAI2017 and CAVAS dataset: approximately 170 subjects from the CAVAS dataset were used for pre-training.

I2I-Mamba: a multi-modal medical image synthesis framework based on selective state space modeling. By leveraging the Mamba architecture, the model captures long-range spatial dependencies with linear computational complexity, enabling efficient modeling of global context for cross-modality translation.

\subsection{Implementation Details}
We implemented our framework using PyTorch. For optimization, we used the Adam optimizer with a learning rate of 2.e{-}4. The beta values for the Adam optimizer were set to (0.5, 0.9) for the discriminator. The batch size was set to 1. For each experiment, our model stabilized after approximately 50 epochs and yielded consistent synthesis results.

The architectures of the generator ($q_{\phi}$) and mutual information estimator ($E$) follow the UNSB framework. The discriminator consists of 6 time-conditional convolution layers~\cite{GUNGOR2023102872}. All experiments were conducted on a single NVIDIA Quadro RTX5000 16GB.

The weight for reconstruction loss, $\lambda_{Rec}$ was set to 0.5 and that for the context-preserving loss, $\lambda_{cpl}$ was set to 10.0. All other weight parameters were set to 1.0. To reduce stochastic output fluctuations, the variance of the random noise was constrained. 

\begin{table}
\renewcommand{\arraystretch}{0.95}
\resizebox{\textwidth}{!}{%
  \begin{tabular}{l l l c c c c c}
  \toprule
  \multirow{2}{*}{Datasets} & \multirow{2}{*}{Task} & \multirow{2}{*}{Methods} & \multicolumn{5}{c}{Metrics} \\
  \cmidrule(lr){4-8}
  & & & PSNR $\uparrow$ & SSIM $\uparrow$ & NRMSE $\downarrow$ & DICE $\uparrow$ & Recall $\uparrow$ \\
  \midrule

  \multirow{5}{*}{IXI} & \multirow{5}{*}{T1w$\rightarrow$T2w} 
   & Pix2Pix & 26.81$\pm$1.42  & 0.914$\pm$0.024 & 0.291$\pm$0.035 & & \\
   & & MT-Net & 26.98$\pm$1.68 & 0.922$\pm$0.019 & 0.295$\pm$0.034  & & \\
   & & DDGAN & 27.66$\pm$1.47 & \textcolor{red}{0.933}$\pm$0.025 & 0.251$\pm$0.057 & & \\
   & & I2I-Mamba & \textcolor{blue}{27.83}$\pm$1.76 & 0.929$\pm$0.025 & \textcolor{red}{0.249}$\pm$0.071 & & \\
   & & Ours & \textcolor{red}{27.97}$\pm$1.53 & \textcolor{blue}{0.929}$\pm$0.022 & \textcolor{blue}{0.251}$\pm$0.052 & & \\
  \midrule 

  \multirow{10}{*}{BraTS2020} & \multirow{5}{*}{T1w$\rightarrow$FLAIR} 
    & Pix2Pix & 23.43$\pm$2.42 & 0.869$\pm$0.021 & 0.287$\pm$0.129 & & \\
   & & MT-Net & 22.28$\pm$1.67 & 0.833$\pm$0.024 & 0.319$\pm$0.107 & & \\
   & & DDGAN & 23.14$\pm$2.38 & 0.885$\pm$0.021 & 0.284$\pm$0.054 & & \\
   & & I2I-Mamba & \textcolor{blue}{24.61}$\pm$2.38 & \textcolor{blue}{0.874}$\pm$0.026 & \textcolor{blue}{0.243}$\pm$0.077 & & \\
   & & Ours & \textcolor{red}{25.72}$\pm$1.58 & \textcolor{red}{0.894}$\pm$0.035 & \textcolor{red}{0.222}$\pm$0.102 & & \\
  \cmidrule(l){2-8}
   & \multirow{5}{*}{T1w$\rightarrow$T2w} 
   & Pix2Pix & \textcolor{red}{25.98}$\pm$2.37 & 0.911$\pm$0.021 & \textcolor{red}{0.251}$\pm$0.054  & & \\
   & & MT-Net & 24.95$\pm$2.37 & 0.916$\pm$0.024 & 0.274$\pm$0.107 & & \\
   & & DDGAN & 24.94$\pm$2.41 & \textcolor{blue}{0.921}$\pm$0.021 & 0.277$\pm$0.054 & & \\
   & & I2I-Mamba & 24.83$\pm$3.91 & 0.915$\pm$0.031 & 0.302$\pm$0.133 & & \\
   & & Ours & \textcolor{blue}{24.98}$\pm$1.53 & \textcolor{red}{0.931}$\pm$0.019 & \textcolor{blue}{0.271}$\pm$0.109 & & \\
   \midrule

  \multirow{5}{*}{\makecell[l]{VALDO}} & \multirow{5}{*}{T1w$\rightarrow$T2*} 
   & Pix2Pix & 26.89$\pm$1.62  & 0.905$\pm$0.025 & \textcolor{blue}{0.211}$\pm$0.058 & & \\
   & & MT-Net & 25.97$\pm$2.16 & 0.885$\pm$0.024 & 0.189$\pm$0.069 & & \\
   & & DDGAN & 24.42$\pm$2.39 & 0.846$\pm$0.069 & 0.258$\pm$0.063 & & \\
   & & I2I-Mamba & \textcolor{blue}{26.89}$\pm$1.53 & \textcolor{blue}{0.906}$\pm$0.025 & 0.219$\pm$0.046 &  &  \\
   & & Ours & \textcolor{red}{28.11}$\pm$1.35 & \textcolor{red}{0.919}$\pm$0.023 & \textcolor{red}{0.184}$\pm$0.041 & &  \\
  \midrule
  
  \multirow{6}{*}{\makecell[l]{MICCAI2017\\WMH}} & \multirow{6}{*}{T1w$\rightarrow$FLAIR} 
  & Pix2Pix & 24.38$\pm$1.91 & 0.891$\pm$0.026 & 0.311$\pm$0.092 & \textcolor{blue}{0.325}$\pm$0.294 & \textcolor{blue}{0.311}$\pm$0.249 \\
   & & MT-Net & 25.21$\pm$2.08 & 0.884$\pm$0.029 & 0.283$\pm$0.089 &  0.1002$\pm$0.05  & 0.092$\pm$0.039 \\
   & & DDGAN & 25.45$\pm$1.95 & 0.899$\pm$0.026 & 0.278$\pm$0.111 & 0.112$\pm$0.087 & 0.102$\pm$0.11 \\
   & & I2I-Mamba & \textcolor{blue}{25.38}$\pm$1.78 & \textcolor{blue}{0.909}$\pm$0.026 & \textcolor{blue}{0.276}$\pm$0.035 & 0.315$\pm$0.226 & 0.274$\pm$0.233 \\
   & & Ours & \textcolor{red}{27.39}$\pm$1.73 & \textcolor{red}{0.928}$\pm$0.021 & \textcolor{red}{0.215}$\pm$0.032 & \textcolor{red}{0.381}$\pm$0.291& \textcolor{red}{0.386}$\pm$0.241 \\
   & & $x_B$ & - & - & - & 0.679$\pm$0.241 & 0.778$\pm$0.245 \\
  \midrule
  
  \multirow{5}{*}{CAVAS dataset} & \multirow{5}{*}{T1w$\rightarrow$FLAIR} 
  & Pix2Pix & 25.53$\pm$1.42 & 0.891$\pm$0.019 & \textcolor{blue}{0.169}$\pm$0.031 & \textcolor{blue}{0.339}$\pm$0.238 & 0.415$\pm$0.331  \\
   & & MT-Net & 25.82$\pm$1.18 & 0.892$\pm$0.016 & 0.183$\pm$0.026 &  0.118$\pm$0.131  & 0.147$\pm$0.203 \\
   & & DDGAN & 24.31$\pm$1.24 & 0.834$\pm$0.018 & 0.208$\pm$0.061 & 0.051$\pm$0.193 & 0.036$\pm$0.091 \\
   & & I2I-Mamba & \textcolor{red}{26.04}$\pm$1.78 & \textcolor{red}{0.911}$\pm$0.024 & \textcolor{red}{0.149}$\pm$0.035 & 0.312$\pm$0.226 & \textcolor{blue}{0.499}$\pm$0.315 \\
   & & Ours & \textcolor{blue}{25.95}$\pm$1.32 & \textcolor{blue}{0.909}$\pm$0.023 & 0.171$\pm$0.018 & \textcolor{red}{0.421}$\pm$0.224 & \textcolor{red}{0.669}$\pm$0.311 \\
  \midrule

  \end{tabular}%
  }
  
\caption{Quantitative Results of the Entire Experiment. \textcolor{red}{Red} indicates the best performance, and \textcolor{blue}{blue} indicates the second-best performance.}
\label{total_result}
\end{table}

\section{Results}
\subsection{Brain MR Image synthesis results}

All experiments used T1w as the source modality, which serves as an optimal source modality for synthesis due to its rapid acquisition time, clinical efficacy, and superior anatomical delineation~\cite{10.1162/IMAG.a.4}.

\subsubsection{IXI dataset (T1w $\rightarrow$ T2w)}
On the IXI dataset, our method achieved the highest PSNR of 
$27.97 \pm 1.53$ dB among all compared methods, surpassing I2I-Mamba 
($27.83 \pm 1.76$), DDGAN ($27.66 \pm 1.47$), and MT-Net 
($26.98 \pm 1.68$). In terms of SSIM, our method ($0.929 \pm 0.022$) 
was competitive with I2I-Mamba ($0.929 \pm 0.025$) and DDGAN 
($0.933 \pm 0.025$), while outperforming Pix2Pix ($0.914 \pm 0.024$) 
and MT-Net ($0.922 \pm 0.019$). The IXI dataset represents a relatively 
standard T1w-to-T2w translation task with normal anatomy and 
no pathological lesions, and the competitive performance of our method 
on this dataset indicates that the Schr\"{o}dinger bridge formulation does 
not compromise general synthesis quality even in the absence of 
lesion-specific priors.

\subsubsection{BraTS2020 dataset (T1w $\rightarrow$ FLAIR, T1w $\rightarrow$ T2w)}
On the BraTS2020 T1w$\rightarrow$FLAIR task, our method achieved the best 
performance across all three metrics, with a PSNR of $25.72 \pm 1.58$ dB, 
SSIM of $0.894 \pm 0.035$, and NRMSE of $0.222 \pm 0.102$, outperforming 
I2I-Mamba ($24.61 \pm 2.38$ / $0.874 \pm 0.026$ / $0.243 \pm 0.077$) 
and DDGAN ($23.14 \pm 2.38$ / $0.885 \pm 0.021$ / $0.284 \pm 0.054$) 
by a substantial margin. Notably, MT-Net showed the weakest performance 
on this task ($22.28 \pm 1.67$ / $0.833 \pm 0.024$ / $0.319 \pm 0.107$), 
suggesting that MAE-based pre-training might not have generalized effectively with limited fine-tuning data. On the T1w$\rightarrow$T2w 
task, our method achieved the highest SSIM ($0.931 \pm 0.019$) while 
maintaining competitive PSNR ($24.98 \pm 1.53$), demonstrating consistent 
structural fidelity across multiple target modalities within the same 
dataset.

\subsubsection{VALDO dataset (T1w $\rightarrow$ T2*)}
The VALDO dataset presented a particularly challenging synthesis scenario, 
as T2* image synthesis from T1w represents a more difficult cross-contrast 
translation task than FLAIR synthesis, and the dataset contains a very 
limited number of axial slices per subject. Despite these challenges, 
our method achieved the best performance across all metrics, with a PSNR 
of $28.11 \pm 1.35$ dB, SSIM of $0.919 \pm 0.023$, and NRMSE of 
$0.184 \pm 0.041$, substantially outperforming all baselines including 
Pix2Pix ($26.89 \pm 1.62$ / $0.905 \pm 0.025$), I2I-Mamba 
($26.89 \pm 1.53$ / $0.906 \pm 0.025$), MT-Net 
($25.97 \pm 2.16$ / $0.885 \pm 0.024$), and DDGAN 
($24.42 \pm 2.39$ / $0.846 \pm 0.069$). This result demonstrates 
the robustness of our framework under severe data scarcity and 
challenging cross-modality translation conditions.

\subsubsection{MICCAI2017 WMH dataset (T1w $\rightarrow$ FLAIR)}
On the MICCAI2017 WMH challenge dataset, which includes ground-truth WMH 
segmentation annotations, our method achieved the best performance across 
all five metrics. In terms of image quality, our method attained a PSNR of 
$27.39 \pm 1.73$ dB and SSIM of $0.928 \pm 0.021$, outperforming all 
baselines by a clear margin. More critically, our method achieved the 
highest WMH Dice score of $0.381 \pm 0.291$ and Recall of 
$0.386 \pm 0.241$, substantially surpassing Pix2Pix 
($0.325 \pm 0.294$ / $0.311 \pm 0.249$) and I2I-Mamba 
($0.315 \pm 0.226$ / $0.274 \pm 0.233$), whereas MT-Net 
($0.100 \pm 0.050$ / $0.092 \pm 0.039$) and DDGAN 
($0.112 \pm 0.087$ / $0.102 \pm 0.110$) failed to reliably preserve 
WMH regions. These results confirm that the incorporation of 
lesion-specific priors via $\mathcal{L}_{cpl}$ and $\mathcal{L}_{wReg}$ 
substantially improves the preservation of spatially sparse, 
clinically critical lesions that are systematically lost by 
competing methods. It is worth noting that the upper bound of WMH Dice 
for the real FLAIR images is $0.679 \pm 0.241$, 
reflecting the inherent difficulty of the task.

\subsubsection{CAVAS dataset (T1w $\rightarrow$ FLAIR)}
On the CAVAS dataset, our method achieved 
the highest WMH Dice score of $0.421 \pm 0.224$ and Recall of 
$0.669 \pm 0.311$, outperforming all baselines. Notably, while I2I-Mamba 
achieved a competitive Recall of $0.499 \pm 0.315$, its Dice score 
($0.312 \pm 0.226$) was substantially lower than ours, suggesting a higher 
rate of false positives. Pix2Pix showed moderate lesion preservation 
($0.339 \pm 0.238$ / $0.415 \pm 0.331$), whereas MT-Net 
($0.118 \pm 0.131$ / $0.147 \pm 0.203$) and DDGAN 
($0.051 \pm 0.193$ / $0.036 \pm 0.091$) largely failed to reconstruct WMH 
regions. In terms of image quality metrics, our method achieved 
competitive PSNR ($25.95 \pm 1.32$) and SSIM ($0.909 \pm 0.023$), 
comparable to the best-performing baseline I2I-Mamba 
($26.04 \pm 1.78$ / $0.911 \pm 0.024$). These results demonstrate 
that our framework generalizes well to real-world clinical data 
beyond publicly available benchmark datasets.

\begin{table}[!h]
\centering
\renewcommand{\arraystretch}{0.65}
\resizebox{\textwidth}{!}{%
\begin{tabular}{lccccc}
\hline
& Pre-trained & Training Params(M) & Inference Params(M) & Training times(s) & Inference times(s) \\
\hline
Pix2Pix & X & 14.14 & 11.37 & 0.31 & 0.12 \\
DDGAN & X & 67.47 & 38.94 & 3.25 & 1.79 \\
MT-Net & O & 139.24 & 137.34 & - & 1.67 \\
I2I-Mamba & X & 112.06 & 109.31 & 0.91 & 0.53 \\
Ours & X & 43.98 & 14.67 & 0.55 & 0.69 \\
\hline
\end{tabular}
}
\caption{Comparison of model complexity and training/inference time.}
\label{tab:complexity}
\end{table}

\begin{figure}[!ht]
\centering
\includegraphics[width=0.75\textwidth]{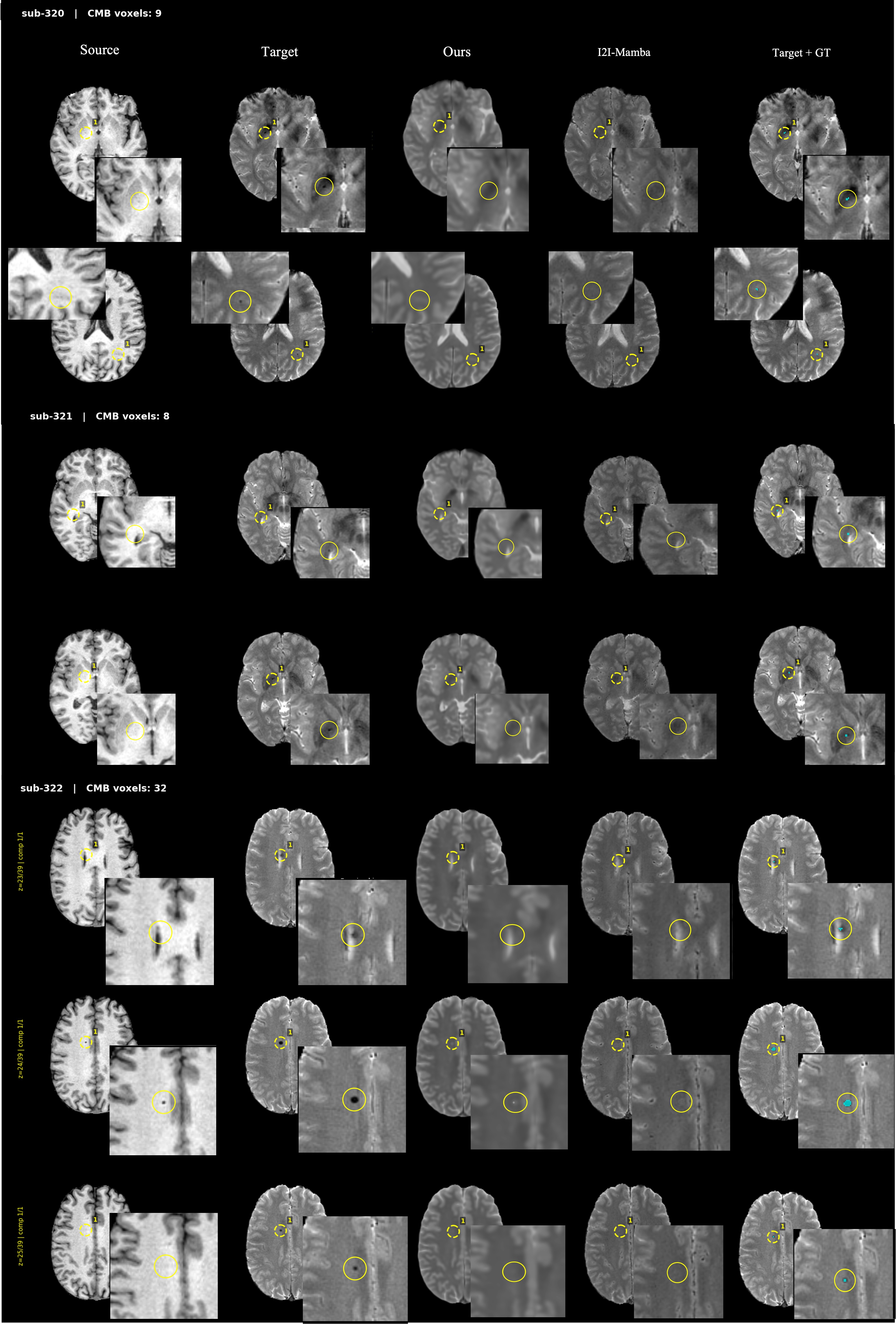}
\caption{Qualitative comparison for CMB slices between our method and other methods.}
\label{VALDO_visualization}
\end{figure}

\subsection{Qualitative results for T2* synthesis}
While our method achieved superior overall image quality in T2* image synthesis, both FGSB and I2I-Mamba failed to reconstruct cerebral microbleeds (CMBs). It should be noted that the CMB evaluation presented here is qualitative, as annotated CMB labels exist for only a very limited number of slices, and the extremely small size and low contrast of CMB lesions make reliable quantitative assessment on synthesized T2* images inherently challenging. We attribute this failure to two primary factors. First, CMB-associated signal characteristics in T1w images are highly ambiguous, as they are easily confounded with real tissue intensity due to the subtle and localized nature of CMB lesions. Second, the slice-wise training paradigm introduces inter-slice inconsistency: while CMB features may be partially visible in T1w images on certain slices, the corresponding signal is absent in immediately adjacent slices, where CMBs are only discernible in T2* images. Under the paired training scheme, this inconsistency causes the model to receive contradictory supervisory signals across neighboring slices, making it difficult to learn a reliable T1w-to-T2* translation for CMB lesions. Figure~\ref{VALDO_visualization} illustrates failure cases across the selected models.

\subsection{Model Complexity Comparison}
Table~\ref{tab:complexity} presents a comparison of model complexity and computational cost across all evaluated methods in terms of trainable parameters and per-iteration training and inference times. Our method did not require any pre-training, in contrast to MT-Net, which depends on MAE-based pre-training using a ViT encoder. While MT-Net's pre-training overhead is not reflected in the per-iteration training time reported here, it constitutes a substantial additional computational burden that must be accounted for in practice, particularly given that effective MAE-based pre-training itself requires a large amount of data. In terms of parameter count, our framework employs 43.98M training parameters, which is substantially lower than MT-Net (139.24M) and I2I-Mamba (112.06M), and comparable to DDGAN (67.47M). Notably, inference requires only 14.67M parameters, as the mutual information estimator $E$, the discriminator $D$, and the patchNCE network $F$ are used exclusively during training. This training-inference parameter asymmetry is a favorable property for deployment, and is more pronounced in our method than in MT-Net (139.24M $\rightarrow$ 137.34M) or I2I-Mamba (112.06M $\rightarrow$ 109.31M), where training and inference parameter counts remained nearly identical. Pix2Pix achieved the smallest parameter count (14.14M training, 11.37M inference) and the fastest training and inference times, but at the cost of significantly reduced synthesis quality and lesion preservation, as evidenced in Table~\ref{total_result}. Regarding per-iteration training time, our method (0.55s) was between Pix2Pix (0.31s) and I2I-Mamba (0.91s), and was substantially lower than DDGAN (3.25s). The higher training cost of DDGAN is attributable to its computationally intensive generator and discriminator networks. At inference, our method required 0.69s per iteration, which is higher than Pix2Pix (0.12s) and I2I-Mamba (0.53s) due to the iterative multi-step generation process inherent to the Schrödinger bridge formulation. However, this cost remains modest and practical, particularly given that the number of function evaluations ($NFE$) is limited to 5 in our experiments. Taken together, our framework achieved a favorable balance between model complexity, training efficiency, and synthesis quality, demonstrating that high-fidelity lesion-preserving synthesis does not necessitate large-scale pre-training or architectures.

\subsection{Ablation Study}
\subsubsection{Component Analysis}
We evaluated the contribution of each loss component on the MICCAI2017 WMH challenge dataset (Table~\ref{tab:ablation}). Basic configuration employed both adversarial loss and the L1 reconstruction loss, serving as the baseline. Incorporating $\mathcal{L}_{SB}$ yielded consistent improvements across all metrics, with PSNR increasing from 26.18 to 27.16 dB, SSIM from 0.901 to 0.919, Dice from 0.287 to 0.299, and Recall from 0.194 to 0.251. These improvements demonstrate that the mutual information-based trajectory regularization effectively stabilizes intermediate sample generation, as intermediate samples $x_{t_i}$ under low-$NFE$ conditions are subject to increasing stochasticity that can cause trajectory drift without explicit regularization. Although the improvement in Dice at this stage is modest, the consistent gains across all metrics confirm that $\mathcal{L}_{SB}$ contributes meaningfully to both image fidelity and lesion preservation.
The addition of lesion-specific losses $\mathcal{L}_{seg}$ (comprising $\mathcal{L}_{cpl}$ and $\mathcal{L}_{wReg}$) yielded the most substantial improvement, particularly in lesion-related metrics. Dice increased markedly from 0.299 to 0.381, and Recall from 0.307 to 0.386, while PSNR and SSIM further improved to 27.39 dB and 0.928, respectively. This result highlights that standard image quality metrics alone are insufficient to capture lesion preservation performance, as the improvement in Dice ($+0.082$) is disproportionately larger than the improvement in PSNR ($+0.23$ dB), reflecting the limited sensitivity of pixel-level metrics to spatially sparse structures such as WMHs. The context-preserving loss $\mathcal{L}_{cpl}$ directly penalizes reconstruction error within lesion regions, while the weighted patchNCE loss $\mathcal{L}_{wReg}$ prioritizes lesion patches during contrastive learning, together providing complementary supervisory signals that enforce clinically relevant lesion preservation.

\begin{table}[ht]
\centering
\renewcommand{\arraystretch}{0.6}
\scalebox{0.65}{%
\resizebox{\textwidth}{!}{%
\begin{tabular}{cccc|cccc}
\toprule
\multicolumn{4}{c|}{Component} & \multicolumn{4}{c}{T1w $\rightarrow$ FLAIR} \\
\cmidrule(r){1-4} \cmidrule(l){5-8}
Basic & $\mathcal{L}_{SB}$ & $\mathcal{L}_{seg}$ & 
& PSNR$\uparrow$ & SSIM$\uparrow$ & Dice$\uparrow$ & Recall$\uparrow$ \\
\midrule
$\checkmark$ &              &              &
& 26.18 & 0.901 & 0.287 & 0.194 \\
 $\checkmark$ & $\checkmark$ &              &
& 27.16 & 0.919 & 0.299 & 0.251 \\
$\checkmark$ & $\checkmark$ & $\checkmark$ &
& \textbf{27.39} & \textbf{0.928} & \textbf{0.381} & \textbf{0.386} \\
\bottomrule
\end{tabular}%
}
}
\caption{Component ablation studies on the MICCAI2017 WMH challenge dataset. $\uparrow$ indicates higher is better.}
\label{tab:ablation}
\end{table}

\begin{table}[ht]
\centering
\renewcommand{\arraystretch}{0.6}
\scalebox{0.85}{%
\begin{tabular}{l|cccc}
\toprule
\multirow{2}{*}{} & \multicolumn{4}{c}{T1w $\rightarrow$ FLAIR} \\
\cmidrule(l){2-5}
& PSNR$\uparrow$ & SSIM$\uparrow$ & Dice$\uparrow$ & Recall$\uparrow$ \\
\midrule
\multirow{2}{*}{FGSB($NFE$=5)}
& 27.39 & 0.928 & \textbf{0.381} & \textbf{0.386} \\
& $\pm$1.73 & $\pm$0.021 & $\pm$0.291 & $\pm$0.241 \\
\midrule
\multirow{2}{*}{FGSB($NFE$=10)}
& 27.15 & 0.917 & 0.352 & 0.344 \\
& $\pm$1.71 & $\pm$0.024 & $\pm$0.279 & $\pm$0.276 \\
\midrule
\multirow{2}{*}{FGSB($NFE$=20)}
& 27.29 & 0.916 & 0.345 & 0.359 \\
& $\pm$1.72 & $\pm$0.022 & $\pm$0.281 & $\pm$0.284 \\
\bottomrule
\end{tabular}%
}
\caption{Ablation study of the iterative refinement process on the MICCAI2017 WMH challenge dataset. Bold values indicate the best performance.}
\label{tab:ablation2}
\end{table}

\subsubsection{Effect of Number of Function Evaluations ($NFE$)}
Table~\ref{tab:ablation2} presents an ablation study on the number of function evaluations ($NFE$), which controls the number of intermediate steps in the generation trajectory. Following the recommendation of the UNSB framework, we evaluated $NFE$ $\in \{5, 10, 20\}$. $NFE$$=$5 achieved the best performance across all metrics, with PSNR of $27.39 \pm 1.73$ dB, SSIM of $0.928 \pm 0.021$, Dice of $0.381 \pm 0.291$, and Recall of $0.386 \pm 0.241$. Increasing $NFE$ to 10 led to a decline across all metrics, with Dice dropping to $0.352 \pm 0.279$ and Recall to $0.344 \pm 0.276$. A further increase to $NFE$$=$20 did not recover performance, yielding Dice of $0.345 \pm 0.281$ and Recall of $0.359 \pm 0.284$. These results confirm that a compact $NFE$ is not merely a computational convenience but a design choice that is empirically optimal within our framework, and that the mutual information regularization $\mathcal{L}_{SB}$ plays a critical role in maintaining trajectory stability under this low-$NFE$ regime.

\section{Discussion and Conclusion}

Despite differences in MRI sequences, imaging protocols, and data acquisition environments, our proposed FGSB framework demonstrates robust 
training capability, achieving reliable image synthesis quality and preserving critical 
lesion features without requiring pre-training procedures. The ablation study 
confirms that each component contributes meaningfully to the final performance, and 
experimental results across multiple datasets validate the generalization performance of the 
framework under diverse data conditions.

A key component of FGSB is the mutual information estimator $E$, which enforces 
trajectory-level consistency by preventing intermediate samples $x_{t_i}$ from 
deviating excessively from the generation trajectory, even when operating under a 
small $NFE$. This regularization is theoretically grounded in the Schr\"{o}dinger bridge 
formulation, where the stochastic interpolation of intermediate states is governed by 
a predefined noise schedule, and maintaining alignment along this trajectory is 
essential for stable convergence. However, the MINE-based estimator~\cite{pmlr-v80-belghazi18a} 
presents a practical limitation: it optimizes a lower bound on mutual information 
rather than the true value, and its adversarial training dynamics are known to be 
sensitive to hyperparameter choices and prone to instability. Consequently, although 
$\mathcal{L}_{SB}$ provides meaningful regularization in our framework, it may not 
represent the optimal solution. Future work may explore alternative mutual 
information estimators or contrastive objectives that offer tighter bounds and more 
stable training behavior.

The current framework operates on 2D axial slices, which facilitates deployment in resource-constrained local environments and enables stable training with a compact model architecture. In several experiments, high-resolution axial images were used, ensuring that sufficient in-plane anatomical detail was preserved for reliable synthesis and lesion evaluation. Nevertheless, the slice-wise paradigm inherently neglects inter-slice consistency, which may limit synthesis quality for structures that extend significantly across multiple slices. This limitation is particularly relevant for images with isotropic voxel sizes. Extending the framework to 3D-based synthesis, with appropriate architectural adaptations to manage computational cost, remains an important direction for future work.

As discussed in Section 5.2, FGSB failed to 
reconstruct CMBs in synthesized T2* images. Consequently, one-to-one synthesis 
from T1w alone is fundamentally limited for CMB-related tasks. To address 
this, future work should explore many-to-one synthesis frameworks that leverage 
multiple source modalities simultaneously. For instance, combining T1w 
with susceptibility-weighted imaging (SWI) or other sequences that carry complementary 
contrast information would provide the model with the necessary input signal 
to support reliable CMB reconstruction. 

Furthermore, several limitations of the current study should be acknowledged. First, the framework 
is designed for paired learning and is therefore not directly applicable to federated 
learning scenarios where only unpaired data from heterogeneous cohorts are available. Second, the complex preprocessing pipeline, which includes co-registration and brain extraction, may constrain practical deployment in clinical workflows.
Third, the current evaluation is limited to brain MR images, and the extension to additional imaging modalities such as PET and CT remains a direction for future investigation.

In this work, we proposed FGSB, a Schr\"{o}dinger bridge-based framework for brain MR 
image synthesis. By incorporating paired target modality 
information and lesion-specific priors into the training stage, 
FGSB achieved high-fidelity synthesis with strong lesion preservation while remaining 
computationally efficient and free from pre-training requirements. The mutual 
information estimator $E$ plays a critical role in stabilizing the generation 
trajectory under low-$NFE$ conditions, and the lesion-specific losses $\mathcal{L}_{cpl}$ 
and $\mathcal{L}_{wReg}$ enable reliable preservation of spatially sparse structures 
such as WMHs that are systematically lost by competing methods. Experimental results 
across five datasets demonstrate that FGSB consistently outperforms competing methods 
on clinically relevant lesion metrics, confirming the effectiveness of 
trajectory-level regularization and prior-guided contrastive learning for medical 
image synthesis.

\section*{Declaration of generative AI and AI-assisted technologies in the writing process.}

During the preparation of this work the authors used Claude (Anthropic) in order to scientific writing. After using this tool, the authors reviewed and edited the content as needed and take full responsibility for the content of the published article.


\section*{Data availability statement}

Publicly available datasets analyzed in this study include VALDO, IXI, MICCAI2017 WMH challenge, and BraTS2020, which can be accessed via their respective official data portals. BraTS 2020 data are available upon registration/data request through the challenge portal. Our code and CAVAS dataset will be made available on request.

\section*{Ethics statement}
IRB information for CAVAS dataset: HY—Hanyang University: HYUIRB-202011-012, CN—Chonnam National University: 06–062, KB—Keimyung University: 2020–01-058, PH—Wonju YonseiUniversity College of Medicine: CR320120, GH—Yonsei University Medical School: 4–2020-0817, Yongin Severance dataset: eIRB-9-2025-0165. The studies were conducted in accordance with the local legislation and institutional requirements. The participants provided their written informed consent to participate in this study.

\section*{CRediT authorship contribution statement}

\textbf{Hanyeol Yang}: Conceptualization, Methodology, Investigation, Software, Validation, Visualization, Writing – original draft, Writing – review and editing. 
\textbf{Sunggyu Kim}: Writing – review and editing, Validation, funding acquisition. 
\textbf{Yongseon Yoo}: Writing – review and editing, Validation. 
\textbf{Mi Kyung Kim}: Resources, Data curation.
\textbf{Yu-Mi Kim}: Resources, Data curation.
\textbf{Min-Ho Shin}: Resources, Data curation.
\textbf{Insung Chung}: Resources, Data curation.
\textbf{Sang Baek Koh}: Resources, Data curation.
\textbf{Jong-Min Lee}: Conceptualization, Writing review and editing, Validation, supervision.
\label{credut}

\section*{Funding}
This work was supported by Institute of Information \& communications Technology Planning \& Evaluation (IITP) grant funded by the Korea government(MSIT) (No.RS-2020-II201373, Artificial Intelligence Graduate School Program(Hanyang University)), This work was supported by the National IT Industry Promotion Agency(NIPA), an agency under the MSIT and with the support of the Daegu Digital Innovation Promotion Agency (DIP), the organization under the Daegu Metropolitan Government. 

\section*{Acknowledgments}
The authors thank the IXI Consortium for making the IXI Dataset publicly available under CC BY-SA 3.0 (https://brain-development.org/ixi-dataset/). This study used the BraTS2020 challenge datasets provided by the organizers of the Multimodal Brain Tumor Segmentation Challenge. The MICCAI 2017 WMH Segmentation Challenge dataset, organized by UMC Utrecht, VU Amsterdam, and NUHS Singapore, was used for training and evaluation. We kindly ask for the following funding sources to be acknowledged: Wellcome Trust (082464/Z/07/Z), British Heart Foundation \\ (SP/07/001/23603, PG/08/103, PG/12/29/29497 and CS/13/1/30327), Erasmus MC University Medical Center, the Erasmus University Rotterdam, the Netherlands Organization for Scientific Research (NWO) Grant 918-46-615, the Netherlands Organization for Health Research and Development (ZonMW), the Research Institute for Disease in the Elderly (RIDE), and the European Union Seventh Framework Programme (FP7/2007–2013) under grant agreement No. 601055, VPHDARE@IT, the Dutch Technology Foundation STW. We thank the VALDO Challenge organizers for providing the data and the ALFA Study contributors for the ALFA Study: Müge Akinci, Annabella Beteta, Raffaele Cacciaglia, Alba Cañas, Irene Cumplido, Carme Deulofeu, Ruth Dominguez, Maria Emilio, Carles Falcón, Karine Fauria, Sherezade Fuentes, Juan Domingo Gispert, Oriol Grau-Rivera, José M. González-de-Echávarri, Laura Hernandez, Gema Huesa, Jordi Huguet, Iva Knezevic, Eider M. Arenaza-Urquijo, Eva M Palacios, Paula Marne, Tania Menchón, Marta Milà-Alomà, Carolina Minguillon, José Luis Molinuevo, Grégory Operto, Albina Polo, Gemma Salvadó, Sandra Pradas, Blanca Rodríguez, Aleix Sala-Vila, Gonzalo Sánchez-Benavides, Mahnaz Shekari, Anna Soteras, Marc Suárez-Calvet, Laura Stankeviciute, Marc Vilanova and Natalia Vilor-Tejedor. We would also like to thank the clinical team of the CAVAS dataset for providing the data resources for this research.



\bibliographystyle{elsarticle-num-names} 
\bibliography{refs}           






\end{document}